\documentstyle[preprint,eqsecnum,aps]{revtex}
 \tightenlines
\draft

\newcommand{\be}{\begin{equation}}
\newcommand{\ee}{\end{equation}}
\newcommand{\beal}{\begin{eqalign}}
\newcommand{\eeal}{\end{eqalign}}
\newcommand{\bea}{\begin{eqnarray}}
\newcommand{\eea}{\end{eqnarray}}
\newcommand{\bean}{\begin{eqnarray*}}
\newcommand{\eean}{\end{eqnarray*}}
\newcommand{\ba}{\begin{array}}
\newcommand{\ea}{\end{array}}

\newcommand{\ka}{\kappa}
\newcommand{\La}{\Lambda}
\newcommand{\la}{\lambda}

\newcommand{\de}{\delta}

\newcommand{\pa}{\partial}

\newcommand{\no}{\nonumber}

\newcommand{\tr}{\mbox{tr}}
\newcommand{\res}{\mbox{res}}

\begin{document}

\title
 { \sc $W_n^{(\ka)}$ algebra associated with the \\
 Moyal KdV Hierarchy\/}
\author{
{\sc Ming-Hsien Tu\footnote{E-mail: phymhtu@ccunix.ccu.edu.tw}\/}\\
  {\it Department of Physics, National Chung Cheng University,\\
   Minghsiung, Chiayi, Taiwan\/}\\
   }
\date{\today}
\maketitle
\begin{abstract}
We consider the Gelfand-Dickey (GD) structure defined by the Moyal $\star$-product with
 parameter $\ka$, which not only defines the bi-Hamiltonian
  structure for the generalized Moyal KdV hierarchy but also provides a $W_n^{(\ka)}$
  algebra containing the Virasoro algebra as a subalgebra with central charge
   $\ka^2(n^3-n)/3$. The free-field realization of the $W_n^{(\ka)}$ algebra is given through
   the Miura transformation and the cases for $W_3^{(\ka)}$ and
  $W_4^{(\ka)}$ are worked out  in detail.
 \end{abstract}

\newpage
%%%%%%%%%%%%%%%%%%%%%%%%%%%%%%%%%%%%%%%%%%%%%%%%%%%%%%%%%%%%%%%%%%%%
\section{Introduction}
Over the past decade, much attention has been paid to the $W$-algebars in connection with
 integrable systems and string theories (see, for example, \cite{BS,BF,Mo} and references therein).
 For example, it is well-known that the classical realizaion of $W_n$
  algebra \cite{Z,B,DIZ} can be constructed  from the second Gelfand-Dickey(GD)
   structure \cite{GD,D} associated with the differential Lax
 operator. Furthermore, the GD bracket with graded commutators was proposed \cite{HN,FR1}
 for supersymmetric pseudo-differential operators and  the classical version
 of extended supersymmetric $W$-algebras was obtained.

 Recently, the so called dispersionless Lax equations were considered \cite{K,KG,Kr,TT},
  which are the quasi-classical limit of the ordinary Lax equations. In such limit,
 the ordinary (pseudo-) differential Lax operators are replaced by the formal
 Laurent series in $p$ and the canonical Poisson bracket
  $\{f(x,p),g(x,p)\}=\pa_pf\pa_xg-\pa_xf\pa_pg$ takes over the role of the commutator
   $[A,B]=AB-BA$ in the ordinary Lax formalism. Moreover, the quasi-classical limit of
    the GD structure can be defined and the associated classical limit of $W$-algebras
     were constructed \cite{FR}. All the mentioned results reveal that the algebraic
      structures associated with the GD bracket are profound.

  More recently, there has been a great deal of interest to study the Moyal deformations of
  the Lax equations \cite{Ku,T,St,G,Ko} in algebraic and/or geometric ways by using the so-called
   Moyal $\star$-product defined by \cite{Gr,M,BFFLS}
    (for a historical survey of $\star$-product, see \cite{Za})
  \be
f\star g=\sum_{s=0}^{\infty}\frac{\ka^s}{s!}\sum_{j=0}^s(-1)^j{s \choose j}(\pa_x^j\pa_p^{s-j}f)
 (\pa_x^{s-j}\pa_p^jg)
 \label{Moyal2}
\ee
 where $f$ and $g$ are two arbitrary functions on the two-dimensional phase space with
 coordinate $(x,p)$. By (\ref{Moyal2}), the Moyal bracket \cite{Gr,M,BFFLS} is defined by
\be
 \{f,g\}_\ka=\frac{f\star g-g\star f}{2\ka}
 \label{Moyalb}
\ee
 that satisfies the following properties (i) $\{f,g\}_\ka=-\{g,f\}_\ka$ (anti-symmetry)
 (ii) $\{af+bg,h\}_\ka=a\{f,h\}_\ka+b\{g,h\}_\ka$ (linearity)
 (iii) $\{f,\{g,h\}_\ka\}_\ka+\{g,\{h,f\}_\ka\}_\ka+\{h,\{f,g\}_\ka\}_\ka=0$ (Jacobi identity).
 Since the Moyal bracket (\ref{Moyalb}) recovers the canonical Poisson bracket in the limit
 $\ka\to 0$, i.e. $\lim_{\ka \to 0}\{f,g\}_\ka=\pa_pf\pa_xg-\pa_xf\pa_pg$, thus it can be
 viewed as the higher-order derivative  generalization of the canonical Poisson bracket.
 Motivated by the works described above and consulting the dispersionless KdV hierarchy \cite{Kr}
  whose Lax flows are defined by the canonical Poisson bracket, the zero-th order
   term in $\ka$ in (\ref{Moyalb}), we attempt to define the the KdV hierarchy by using the Moyal
   bracket. We shall show that the integrability is still maintained under the
    Moyal deformation and a $W$-type algebra emerges naturally from the associated
     second GD structure.

Our paper is organized as follows: In Sec. II, we recall some basic notions and introduce
 the Lax formulation of the Moyal KdV hierarchy. In Sec. III, we show that the Gelfand-Dickey
 structure with respect to the Moyal $\star$-product defines the bi-Hamiltonian structure of the
 Moyal KdV hierarchy. In Sec. IV, we study the conformal property of the Gelfand-Dickey algebras
 and explicitly identify the first few primary fields. In Sec. V, by factorizing the Lax operator,
 the free-field realization for the associated conformal algebras is given.
 We work out some examples in Sec. VI and present the concluding remarks in Sec. VII.
%%%%%%%%%%%%%%%%%%%%%%%%%%%%%%%%%%%%%%%%%%%%%%%%%%%%%%%%%%%%
\section{Moyal KdV Hierarchy}
 To begin with, let us consider an algebra of Laurent series
 of the form $\La=\{A|A=\sum_{i=-\infty}^Na_ip^i\}$ with coefficients $a_i$ depending
  on an infinite set of variables $t_1\equiv x, t_2, t_3, \cdots$. The algebra $\La$
  can be decomposed into the sub-algebras as $\La=\La_{\ge k}\oplus \La_{< k}, (k=0,1,2.)$
 where $\La_{\ge k}=\{A\in \La| A=\sum_{i\ge k}a_ip^i\},~\La_{< k}=\{A\in\La|
  A=\sum_{i< k}a_ip^i\}$ and using the notations : $\La_+=\La_{\ge 0}$ and $\La_-=\La_{< 0}$
   for short. It's obvious that $\La$ is an  associative but noncommutative algebra under the
 Moyal $\star$-product. For a given Laurent series $A$ we define its residue as $\res(A)=a_{-1}$
 and its trace as $\tr(A)=\int \res(A)$. For any two Laurent series $A=\sum_{i}a_ip^i$ and
$B=\sum_{j}b_jp^{-j}$ we have
 \be
  \int\res(A\star B)=\int\sum_{i,j} \frac{\ka^{i-j+1}i!}{(i-j+1)!(j-1)!}(a_ib_j)^{(i-j+1)}
  =\sum_i\int a_ib_{i+1}
 \label{res}
 \ee
 which  is the same as the case in the dispersionless limit $\ka\to 0$. We shall see that
  it's because of the nice property (\ref{res}) so that the Hamiltonian formulation for
   the Moyal KdV becomes possible.  Using (\ref{res}) it is easy to show that
 $\tr \{A, B\}_\ka=0$ and $\tr (A\star\{B, C\}_\ka)=\tr (\{A, B\}_\ka\star C)$.
 Here we simply remark that, due to the property (\ref{res}), the Moyal $\star$-product
  within the trace can be replaced by the ordinary multiplication.
  However we shall reserve the product for convenience.

  Finally, given a functional $F(A)=\int
f(a)$ we define its gradient as
 \[
d_AF=\sum_i\frac{\de f}{\de a_i}p^{-i-1}
 \]
where the variational derivative is defined by
 \[
\frac{\de f}{\de a_k}=\sum_i(-1)^i\left(\pa^i\cdot\frac{\pa f}{\pa a_k^{(i)}}\right),
 \]
with $a_k^{(i)}\equiv (\pa^i\cdot a_k), \pa\equiv \pa/\pa x$.
 Note that we shall use the notations $\pa\cdot f=f'=\pa f/\pa x$
and $\pa f=f\pa+f'$ in the following sections.

The Moyal KdV hierarchy is defined by the Lax equations
 \bea
  \frac{\pa L}{\pa t_k}&=&\{(L^{1/n})^k_+,L\}_{\ka},\qquad
 (L^{1/n})^k_+=(\underbrace{L^{1/n}\star L^{1/n}\star\cdots \star L^{1/n}}_{k})_+\no\\
 &=&\{L,(L^{1/n})^k_-\}_\ka
 \label{laxeq}
\eea
 where the Lax operator $L=p^n+\sum_{i=0}^{n-1}u_ip^{i}$ is a polynomial in $p$ and
  $L^{1/n}=p+\sum_{i=0}a_ip^{-i}$ is   the $n$th root of $L$ in such a way that
\[
L=\underbrace{L^{1/n}\star L^{1/n}\star\cdots \star L^{1/n}}_{n}.
\]
 From the definition of the Moyal bracket, the highest order in $p$ on the right-hand side
  of the Lax equations (\ref{laxeq}) is $n-2$.
   That means  $u_{n-1}$ is trivial in evolution equations and
  thus can be dropped in the Lax formulation. However, this is not
   the case for the Hamiltonian formulation (see next section).

 Let us work out the simplest example. For $n=2$, $L=p^2+u$ and we have
  $L^{1/2}=p+\sum_{i=1}^\infty a_ip^{-i}$
 with
 \bean
 a_1&=& \frac{1}{2}u,\\
 a_3&=& -\frac{1}{8}u^2,\\
 a_5&=& \frac{1}{16}u^3+\frac{1}{8}\ka^2(u_x^2-2uu_{xx}),\\
 a_7&=& -\frac{5}{128}u^4-\frac{5}{16}\ka^2(uu_x^2-2u^2u_{xx})-
 \frac{1}{8}\ka^4(u_{xx}^2-2u_xu_{xxx}+2uu^{(4)}),\\
 a_9&=& \frac{7}{256}u^5+\frac{35}{64}\ka^2(u^2u_x^2-2u^3u_{xx})+
 \frac{7}{16}\ka^4(uu_{xx}^2-4uu_xu_{xxx}+3u^2u^{(4)}-3u_{xx}u_x^2)\\
 &&+\frac{1}{4}\ka^6(u_xu^{(5)}-uu^{(6)}),
 \eean
 and $a_{2k}=0$ etc. The first few Lax flows are given by
 \bea
 u_{t_1}&=&u_x,\no\\
 u_{t_3}&=&\frac{3}{2}uu_x+\ka^2u_{xxx},\no\\
 u_{t_5}&=&\frac{15}{8}u^2u_x+\frac{5}{2}\ka^2(uu_{xxx}+2u_xu_{xx})+\ka^4u^{(5)},
 \label{Moyalkdv}\\
 &\cdots&\no
  \eea
 The set of equations (\ref{Moyalkdv}) form what we call the Moyal KdV hierarchy which
 can also be obtained from the reduction of the Moyal KP hierarchy \cite{St} or noncommutative
 zero-curvature equations \cite{Ko}. Note that, when $\ka=0$, all higher-order derivative
  terms disappear and the Moyal KdV hierarchy reduces to the dispersionless KdV hierarchy
   which is of the hydrodynamic type \cite{DN}.
   In this sense, the Moyal parameter $\ka$ characterizes the dispersion effect.
  On the other hand, when $\ka=1/2$, the Moyal KdV hierarchy (\ref{Moyalkdv}) recovers
   the ordinary KdV hierarchy. This is not an accident due to the fact that, at $\ka=1/2$,
   the Moyal KP hierarchy is isomorphic with the ordinary KP hierarchy \cite{St,G}. Thus
    the  Moyal KdV is isomorphic with the ordinary KdV as well.
   For example, it is not hard to show that the  Moyal Boussinesq hierarchy ($L=p^3+u_1p+u_0$)
   at $\ka=1/2$ is isomorphic with the ordinary Boussinesq hierarchy ($L=\pa^3+v_1\pa+v_0$)
    by identifying
  \be
 u_1=v_1,\qquad u_0=v_0-\frac{1}{2}v_1'.
 \label{iso}
  \ee
Finally we like to remark that a similar construction was established in \cite{Ku}. However,
 it was pointed out in \cite{St} that the Lax equations obtained there do not have
 dispersionless limit as $\ka\to 0$, not without a scaling transformation.
%%%%%%%%%%%%%%%%%%%%%%%%%%%%%%%%%%%%%%%%%%%%%%%%%%%%%%%%%%%%%%%%%%%%%%%%%%%%%%%
 \section{Bi-Hamiltonian structure}

Having introduced the Lax formalism, in this section, we would  like to discuss the
 Hamiltonian formalism of the Moyal KdV hierarchy.
  For the Lax operator $L=p^n+\sum_{i=0}^{n-1}u_ip^i$ and
   functionals $F[L]$ and $G[L]$ we define the second Gelfand-Dickey bracket \cite{D}
    with respect to the Moyal $\star$-product as
  \be
\{F,G\}_2=\tr(J^{(2)}(d_LF)\star d_LG)=\int \res(J^{(2)}(d_LF)\star d_LG)
 \label{gdb}
  \ee
 in which $J^{(2)}$ is the Adler map defined by \cite{A}
   \bea
  J^{(2)}(X)&=&\{L,X\}_{\ka+}\star L-\{L,(X\star L)_+\}_\ka,\no\\
  &=&\{L,(X\star L)_-\}_\ka-\{L,X\}_{\ka-}\star L
  \label{adler}
  \eea
  where $X=\sum_{i=1}^nx_ip^{-i-1}$. To verify that $J^{(2)}$ is indeed Hamiltonian,
  we have to check that the Poisson bracket defined in (\ref{gdb}) is antisymmetric
  and obeys the Jacobi identity. For antisymmetry, it can be easily shown that
  $\{F,G\}_2=-\{G,F\}_2$ by using the cyclic property of the trace.
  For the Jacobi identity, instead of direct computation, we shall justify it
   by the Kupershmidt-Wilson (KW) theorem \cite{KW} that will be done in Sec.V.

   Since $J^{(2)}(X)$ is linear in $X$ and
   has order at most $n-1$ thus
\[
J^{(2)}(X)=\sum_{i,j=0}^{n-1}(J^{(2)}_{ij}\cdot x_j)p^{i}
\]
where $J^{(2)}_{ij}$ are differential operators.

 To impose the reduction $u_{n-1}=0$,
 the standard Dirac procedure \cite{DIZ} gives
 \be
 \hat{J}^{(2)}(X)=\{L,X\}_{\ka+}\star L-\{L,(X\star L)_+\}_\ka+
 \frac{1}{n}\{L,\int^x\res\{L,X\}_\ka\}_\ka
 \label{rgd}
 \ee
 or, in components,
 \be
\hat{J}^{(2)}_{ij}=J^{(2)}_{ij}-J^{(2)}_{i,n-1}(J^{(2)}_{n-1,n-1})^{-1}J^{(2)}_{n-1,j}.
 \label{dirac}
 \ee
 Hence the reduced Poisson brackets for $u_i$ can  be expressed as
\be
\{u_i(x),u_j(y)\}^D_2=\hat{J}^{(2)}_{ij}\cdot\delta(x-y).
 \label{j2b}
 \ee
  So far we only discuss the second Poisson structure.
 To obtain the first structure, as usual, one can deform the second structure by
 shifting $L\to L+\lambda$ (or $u_0\to u_0+\la$) and extract the first structure
  from the term proportional to $\la$. It turns out that
 \[
 J^{(1)}(X)=\{L,X\}_{\ka+}=\sum_{i=0}^{n-2}(J^{(1)}_{ij}\cdot x_j)p^{i}
 \]
 which is compatible with the reduction $u_{n-1}=0$. The first Poisson brackets for $u_i$
 read
 \be
\{u_i(x),u_j(y)\}_1=J^{(1)}_{ij}\cdot\delta(x-y).
 \label{j1b}
 \ee
Using the GD brackets (\ref{j2b}) and (\ref{j1b}), the Lax flows (\ref{laxeq}) can be written as
 Hamiltonian flows as
 \[
\frac{\pa L}{\pa t_k}=\{H_k,L\}_2^D=\{H_{k+n},L\}_1
 \]
or, in components,
\[
\frac{\pa u_i}{\pa t_k}=\hat{J}_{ij}^{(2)}\cdot\frac{\de H_{k}}{\de u_j}=
 J_{ij}^{(1)}\cdot\frac{\de H_{k+n}}{\de u_j}
\]
where the Hamiltonians $H_k$ are defined by
 \be
  H_k=\frac{n}{k}\int \res(L^{1/n}\star\cdots \star L^{1/n}).
\label{ham}
 \ee
 For $n=2$, $L=p^2+u$ and the bi-Hamiltonian  structure is given by
 \bea
\{u(x),u(y)\}_1&=&2\pa\cdot\delta(x-y),\no\\
 \{u(x),u(y)\}^D_2&=&[2\ka^2\pa^3+2u\pa+u_x]\cdot\delta(x-y).
 \label{kdvb}
 \eea
The first few Hamiltonians from (\ref{ham}) are
 \bean
 H_1&=& \int u,\\
 H_3&=& \frac{1}{4}\int u^2,\\
 H_5&=& \frac{1}{8}\int (u^3+2\ka^2 uu_{xx}),\\
 H_7&=& \frac{1}{64}\int (5u^4-40\ka^2 uu_x^2+16\ka^4u_{xx}^2),
\eean
 which together with (\ref{kdvb}) implies
 \[
 \frac{\pa u}{\pa t_{2n+1}}=[2\ka^2\pa^3+2u\pa+u_x]\cdot \frac{\de H_{2n+1}}{\de u}
 =2\pa\cdot \frac{\de H_{2n+3}}{\de u}.
 \]
 Therefore we obtain the following recursion relation
\[
\frac{\pa u}{\pa t_{2n+1}}=R\cdot\frac{\pa u}{\pa t_{2n-1}}
\]
with the recursion operator \cite{Ko}
\[
R=\hat{J}^{(2)}(J^{(1)})^{-1}=\ka^2\pa+u+\frac{1}{2}u_x\pa^{-1}
\]
where the inverse operator $\pa^{-1}$ is realized as $\pa^{-1}\cdot f=\int^xf$.
%%%%%%%%%%%%%%%%%%%%%%%%%%%%%%%%%%%%%%%%%%%%%%%%%%%%%%%%%%%%%%%%%%%%%
\section{Moyal deformation of classical $W_n$-algebras}
In general, for the $n$th-order generalized Moyal KdV, we can substitute $L$ and $X$
 into $J^{(1)}(X)$ and $\hat{J}^{(2)}(X)$ to read off the Hamiltonian
   operators $J^{(1)}_{ij}$ and $\hat{J}^{(2)}_{ij}$. For the first structure,
    it is quite easy to show
 \bean
J^{(1)}_{ij}&=&n\pa,\qquad (i+j=n-2)\\
 J^{(1)}_{ij}&=&\ka^{<n-j-i-2>}{n \choose <n-j-i-2>+1}\pa^{<n-j-i-2>+1}\\
 & &+\sum_{l=i}^{n-4-j}\sum_{m=0}^{<l-i>+1}\ka^{<l-i>}{l+j+2 \choose <l-i>+1-m}
 {j+m \choose j}u_{l+j+2}^{(m)}\pa^{<l-i>+1-m},\\
&&(i+j\leq n-4)
 \eean
 and $J^{(1)}_{ij}=0$ otherwise, where $<..>=0$ unless it's an even number.
 On the other hand, the case for the second structure is more complicated.
 From (\ref{adler}) we have
\bean
 J^{(2)}(X)&=&\sum_{l=0}^{n-1}\sum_{m=0}^{n-l-1}\sum_{k=0}^{n}\sum_{s=0}^{m+k}\sum_{i=0}^{m+k-s}
 \sum_{q=m}^{n-l-1}\sum_{j=0}^{q-m}\ka^{<k+q-s-1>}(-1)^i{m\choose s-k+i}{k\choose i}\\
 &&{q+l+1\choose j}{q+l-m-j\choose l}u_k^{(m+k-s-i)}
 \left(u_{q+l+1}^{(q-m-j)}x_l^{(j)}\right)^{(i)}p^s.
  \eean
In particular,
 \bean
 J^{(2)}_{n-1,n-1}&=&-n\pa,\\
 J^{(2)}_{s,n-1}&=&\sum_{k=s}^n\ka^{<k-s-1>}(-1)^{k-s}{k\choose s}u_k\pa^{k-s},\\
 J^{(2)}_{s,n-2}&=&\sum_{k=s}^n\sum_{q=0}^1\sum_{j=0}^q\ka^{<k-s+q-1>}(-1)^{k-s}{k\choose s}
 {q+n-1\choose j}{q-j+n-2\choose n-2}u_k\pa^{k-s}u_{q+n-1}^{(q-j)}\pa^j\\
 &&+\sum_{k=s}^n\sum_{i=0}^{1+k-s}\ka^{<k-s>}(-1)^i{1\choose 1+k-s-i}
 {k\choose i}u_k^{1+k-s-i}\pa^i
 \eean
which together with (\ref{dirac}) yields the reduced Poisson algebras
 \bea
  \{u_{n-2}(x),u_{n-2}(y)\}_2^D&=&[\ka^2\frac{(n^3-n)}{3}\pa^3+2u_{n-2}\pa+u'_{n-2}]
  \cdot\de(x-y),\no\\
 \{u_{n-3}(x),u_{n-2}(y)\}_2^D&=&[3u_{n-3}\pa+u'_{n-3}]\cdot\de(x-y),\no\\
 \{u_{n-4}(x),u_{n-2}(y)\}_2^D&=&\left[\ka^4\frac{(n+1)n(n-1)(n-2)(n-3)}{30}\pa^5\right.\no\\
 &&+\ka^2\left(\frac{(n-2)(n-3)(n+2)}{3}u_{n-2}\pa^3+
 \frac{(n-2)(n-3)}{2}u'_{n-2}\pa^2\right)\no\\
 &&\left.+4u_{n-4}\pa+u'_{n-4}\right]\cdot\de(x-y),\no\\
 \{u_{n-3}(x),u_{n-3}(y)\}_2^D&=&\left[-\ka^4\frac{(n+2)(n+1)n(n-1)(n-2)}{45}\pa^5-\ka^2
 \left(\frac{2(n-2)(n+2)}{3}u_{n-2}\pa^3\right.\right.\no\\
 & &\left.+(n-2)(n+2)u'_{n-2}\pa^2+n(n-2)u''_{n-2}\pa+\frac{(n-1)(n-2)}{3}u'''_{n-2}\right)\no\\
 && \left.-\frac{2(n-2)}{n}u_{n-2}\pa u_{n-2}+4u_{n-4}\pa+2u'_{n-4}\right]\cdot\de(x-y).
 \label{pb2}
 \eea
 Thus $u_{n-2}(x)$ can be identified as the energy-momentum $T(x)$ of the Virasoro algebra
  with central charge $c(n,\ka)=\ka^2(n^3-n)/3$ whereas the other coefficient functions $u_i$,
   with respect to $u_{n-2}$, are not primaries except  $u_{n-3}$ which is a spin-3
    primary field $W_3$. However, if we appropriately linearly combine the differential
     polynomials of $u_i$, then the primary fields $W_i (i\geq 4)$ can be constructed such that
 $\{W_i(x),u_{n-2}(y)\}_2^D=[iW_i\pa+W'_i]\cdot\de(x-y)$.
  In summary, we have
 \bea
 T&=&u_{n-2},\no\\
 W_3&=&u_{n-3},
 \label{pri}\\
 W_4&=&u_{n-4}-\ka^2\frac{(n-2)(n-3)}{10}u''_{n-2}-\ka^4
 \frac{(n-2)(n-3)(5n+7)}{10(n^3-n)}u_{n-2}^2\no
  \eea
etc. Based on the above discussions and some computations we expect that the
 $W_n^{(1/2)}$ algebra coincides with the classical $W_n$ algebra associated with the
  ordinary differential Lax operator $L=\pa^n+\sum_{u=0}^{n-2}v_i\pa^i$.
  We remark that the above relationship between the primaries $W_i$ and the coefficient
   functions $u_i$ is different from those tabulated in \cite{DIZ} due to the fact that
 the Moyal KdV  at $\ka=1/2$  is isomorphic with the ordinary KdV.
  As an illustration, again, let us consider the differential Lax operator
   $L=\pa^3+v_1\pa+v_0~(n=3)$. The primary fields are given by \cite{B}
   \[
   T=v_1,\qquad W_3=v_0-\frac{1}{2}v_1'
   \]
 which together with the isomorphism (\ref{iso}) implies
 \[
 T=u_1,\qquad W_3=u_0.
 \]
 as desired. For higher primaries, such as $W_4$ in (\ref{pri}), one can verify the
 corresponding modifications in the same manner.

 Before ending this section, it should be mentioned that, in the limit $\ka\to 0$,
  both algebraic structures $J_{ij}^{(1)}$ and $\hat{J}_{ij}^{(2)}$ will reduce to
   their corresponding dispersionless counterparts \cite{FR}.
   Particularly, for the second structure $\hat{J}_{ij}^{(2)}$, the coefficient functions
    $u_i$ under this limit are already primary fields with respect to $u_{n-2}$,
     the generator of Diff$S^1$.

%%%%%%%%%%%%%%%%%%%%%%%%%%%%%%%%%%%%%%%%%%%%%%%%%%%%%%%%%%%%%%%%%%%%%
\section{KW Theorem and free-field realization}
 Recall that the Poisson bracket defined by the Adler map (\ref{adler}) is given by
 \[
 \{F,G\}_2(L)=\tr\left[J^{(2)}(d_LF)\star d_LG\right]
 \]
 where $F$ and $G$ are functionals of $L$. Suppose the Lax operator $L$ can be
  factorized  as $L=L_1\star L_2$, then from the variation
 \bean
  \de F&=&\tr(d_LF\de L)=\tr(d_{L_1}F\de L_1+d_{L_2}F\de L_2),\\
  &=&\tr(d_{L}F\de L_1\star L_2+d_{L}FL_1\star\de L_2)
 \eean
 and (\ref{res}) we have
\be
d_{L_1}F=L_2d_LF,\qquad d_{L_2}F=d_LFL_1.
 \label{var}
 \ee
  With the use of (\ref{res}) and (\ref{var}) the proof for the KW theorem is almost
   a repetition of that for the dispersionless case \cite{MR}.
    We shall not spell it out here. It turns out that
\[
\{F,G\}_2(L_1\star L_2)=\{F,G\}_2(L_1)+\{F,G\}_2(L_2).
\]
Hence, if we factorize the Lax operator as $L=p^n+\sum_{i=1}^nu_ip^i=
 (p+\phi_1)\star\cdots \star (p+\phi_n)$ which defines a Miura transformation between
  the coefficients $\{u_i\}$ and the Miura variables $\{\phi_i\}$, then the second GD bracket
  becomes
\[
\{F,G\}_2(L)=\sum_{i=1}^n\int (\frac{\de F}{\de \phi_i})'(\frac{\de G}{\de \phi_i})
\]
which yields the Poisson brackets
\[
\{\phi_i(x),\phi_j(y)\}_2=-\de_{ij}\pa\cdot\de (x-y),\qquad i,j=1,\cdots,n
\]
 The above  brackets immediately justify the Jacobi identity for the second
 GD bracket (\ref{gdb}) as we claimed in Sec. III.
To impose the reduction $u_{n-1}=0$, in view of $u_{n-1}=\sum_{i=1}^n\phi_i$, we have
 \be
 \{\phi_i(x),\phi_j(y)\}^D_2=\left[\frac{1}{n}-\de_{ij}\right]\pa\cdot\de (x-y).
 \label{free}
 \ee
 Equation (\ref{free}) enables us to write down the reduced Poisson brackets of $u_i$
 through the Miura transformation.
Since the Poisson matrix in (\ref{free}) is a $n\times n$ symmetric matrix, thus it can be
 diagonalized to obtain the free-field realization of the $W_n^{(\ka)}$ algebra.

%%%%%%%%%%%%%%%%%%%%%%%%%%%%%%%%%%%%%%%%%%%%%%%%%%%%%%%%%%%%%%%%%%%%%%%%%
\section{$W_3^{(\ka)}$ and $W_4^{(\ka)}$-algebras}
 Let us work out the examples $n=3$ and $n=4$. For $n=3$, $L=p^3+u_1p+u_0$ and
  $X=x_1p^{-2}+x_0p^{-1}$ then the Poisson algebras for $u_1=T$ and $u_0=W_3$ are
 \bean
 \{T(x),T(y)\}_2^D&=&[8\ka^2\pa^3+2T\pa+T']\cdot\delta(x-y),\\
 \{W_3(x),T(y)\}_2^D&=&[3W_3\pa+W_3']\cdot\delta(x-y),\\
 \{W_3(x),W_3(y)\}_2^D&=&[-\frac{8}{3}\ka^4\pa^5-\frac{1}{3}\ka^2
 (10T\pa^3+15T'\pa^2+9T''\pa+2T''')-\frac{2}{3}T\pa T]\cdot\delta(x-y)
 \eean
which is $W_3^{(\ka)}$. In the limit $\ka \to 0$, the $W_3^{(0)}$ algebra
 is just $\mbox{w}_3$ \cite{FR} which is a nonlinear extension  of Diff$S^1$ by a spin-3
  primary field, while $\ka=1/2$, $W_3^{(1/2)}$ is nothing but the classical realization of the
 Zamolodchikov's $W_3$ algebra presented in \cite{B}.

 If we factorize the Lax operator as
 \[
 L=(p+\phi_1)\star(p+\phi_2)\star(p-\phi_1-\phi_2)
 \]
then the primary fields can be expressed in terms of the Miura variables as
 \bean
 T&=&-(\phi_1^2+\phi_1\phi_2+\phi_2^2)-2\ka(2\phi_1'+\phi_2'),\\
 W_3&=&-\phi_1\phi_2(\phi_1+\phi_2)-\ka(\phi_2\phi_1'+2\phi_2\phi_2'+3\phi_1\phi_2')
 -2\ka^2\phi_2''.
 \eean
 Through the above Miura transformation the $W_3^{(\ka)}$ can be rederived by using those
 brackets of $\phi_i$.

 For $n=4$, $L=p^4+u_2p^2+u_1p+u_0$ and $X=x_2p^{-3}+x_1p^{-2}+x_0p^{-1}$. Then from (\ref{rgd})
 we have
 \bean
\{u_2(x),u_2(y)\}_2^D&=&[20\ka^2\pa^3+2u_2\pa+u_2']\cdot\de(x-y),\\
 \{u_1(x),u_2(y)\}_2^D&=&[3u_1\pa+u_1']\cdot\de(x-y),\\
 \{u_0(x),u_2(y)\}_2^D&=&[4\ka^4\pa^5+\ka^2(4u_2\pa^3+u'_2\pa^2)+4u_0\pa+u'_0]\cdot\de(x-y),\\
 \{u_1(x),u_1(y)\}_2^D&=&[-16\ka^4\pa^5-2\ka^2(4u_2\pa^3+6u'_2\pa^2+4u''_2\pa+u'''_2)\\
 &&-u_2\pa u_2+4u_0\pa+2u'_0]\cdot\de(x-y),\\
 \{u_0(x),u_1(y)\}_2^D&=&\left[-\ka^2(5u_1\pa^3+4u'_1\pa^2+u''_1\pa)-
 \frac{1}{2}u_1\pa u_2\right]\cdot\de(x-y),\\
 \{u_0(x),u_0(y)\}_2^D&=&[4\ka^6\pa^7+\ka^4(6u_2\pa^5+15u_2'\pa^4+20u_2''\pa^3+
 15u'''_2\pa^2+6u^{(4)}_2\pa+u^{(5)}_2)\\
 &&+\ka^2((4u_0+2u_2^2)\pa^3+(6u_0'+6u_2u_2')\pa^2+(4u_2u_2''+2(u_2')^2+4u_0'')\pa\\
 &&+(u_2u_2'''+u_2'u_2''+u_0'''))+(2u_0u_2-3u_1^2/4)\pa\\
 &&+(u_2u_0'+u_0u_2'-3u_1u_1'/4)]\cdot\de(x-y).
 \eean
 According to (\ref{pri}), if we define $T=u_2, W_3=u_1$, and
 $W_4=u_0-\ka^2u''_2/5-9\ka^4u_2^2/100$ then the brackets involving $W_4$ are
 \bean
 \{W_4(x),T(y)\}_2^D&=&[4W_4\pa+W'_4]\cdot\de(x-y),\\
 \{W_4(x),W_3(y)\}_2^D&=&\left[-\frac{9}{50}\ka^4(3TW_3\pa+2TW'_3)-\frac{2}{5}\ka^2(14W_3\pa^3
  +14W'_3\pa^2+6W''_3\pa+W'''_3)\right.\\
  &&\left. -\frac{1}{2}(W_3T\pa+W_3T')\right]\cdot\de(x-y),\\
 \{W_4(x),W_4(y)\}_2^D&=&\left[\frac{\ka^{10}}{125}(81T^2\pa^3+243TT'\pa^2+
 243TT''\pa+81TT''')\right.\\
 &&-\frac{\ka^8}{2500}(243T^2T'+81T^3\pa)+\frac{\ka^6}{250}(800\pa^7-288T^2\pa^3-864TT'\pa^2\\
 &&+(54(T')^2-918TT'')\pa+27T'T''-315TT''')\\
 &&+\frac{\ka^4}{100}(448T\pa^5+1120T'\pa^4+1344T''\pa^3+896T'''\pa^2\\
 &&+(18T^3+120T^{(4)}-144TW_4)\pa+27T^2T'-72(TW_4)'+48T^{(5)})\\
 &&+\frac{\ka^2}{5}((12W_4+10T^2)\pa^3+(18W_4'+30TT')\pa^2\\
 &&+(10W''_4+22TT''+10(T')^2
 )\pa+2W_4'''+6T'T''+6TT''')\\
 &&\left.+T\pa W_4+W_4\pa T-\frac{3}{4}W_3\pa W_3\right]\cdot\de(x-y)
 \eean
 which together with the rest of the bracket constitute the $W_4^{(\ka)}$.
 Since the discussions for the limiting cases $W_4^{(0)}$ and $W_4^{(1/2)}$ are similar
  to those of $W_3^{(\ka)}$,  hence we omit it here.

 %%%%%%%%%%%%%%%%%%%%%%%%%%%%%%%%%%%%%%%%%%%%%%%%%%%%%%%%%%%%%%%
\section{Concluding remarks}
We have studied the integrability of the Moyal KdV hierarchy from Lax and Hamiltonian
 point of views. We show that the bi-Hamiltonian structure is encoded by the GD structure
 defined by the Moyal $\star$-product. We work out the GD brackets and investigate the associated
 conformal algebras. The $W_n^{(\ka)}$ algebra we obtained can be viewed as an
 one parameter deformation of the classical $W_n$ algebra arising from the differential Lax formalism.
 We prove the KW theorem for the second GD structure by factorizing the Lax operator
 and obtain a $\ka$-independent free-field realization.

 In spite of the results obtained, few remarks are in order. First, we may generalize the
 construction in this paper to the Moyal KP hierarchy \cite{St} and investigate
 its associated $W_{KP}^{(\ka)}$ algebra  by considering the Lax operator
  $L=p^n+\sum_{i=1}^{\infty}u_{n-i}p^{n-i}$.
 We expect that $W_{KP}^{(1/2)}$ would coincide with the $W_{KP}$ algebra \cite{FMR} defined by
 the pseudo-differential operator. Secondly, it would be interesting to search for
 some topological models so that their  RG flows in  module spaces are governed by
 the Moyal KdV equations. Works in these directions are now in progress.

{\bf Acknowledgments\/}\\
 I would like to thank Prof. C. Zachos for his historical comment on the $\star$-product.
  M.H.T thanks the National Science Council of Taiwan (Grant No. NSC 89-2112-M194-020)
   for support.

   Note added: After submission of this manuscript for publication we become
    aware of the preprint by A. Das and Z. Popowicz \cite{DP} which partly overlaps
    our work.

\end{document}